**Blocking of iron magnetic moments and Spin density wave in $FeTi_2S_4$ and $Fe_2TiS_4$**

B. P. Embaid, O.M. Barrios and M.V. Bello

Laboratorio de Magnetismo, Escuela de Física, Universidad Central de Venezuela, Apartado 47586, Los Chaguaramos, Caracas 1041-A, Venezuela.

## Abstract

Iron-titanium sulfides $FeTi_2S_4$ and $Fe_2TiS_4$ have been structurally and magnetically characterized using powder X Ray Diffraction with Rietveld refinement method and $^{57}Fe$ Mössbauer spectroscopy at variable temperature. Both sulfides have the same crystallographic phase, based on the monoclinic $Cr_3S_4$ type structure and vary in atomic coordinates; $FeTi_2S_4$ retains the ideal atomic positions proposed for the $Cr_3S_4$ phase, while in $Fe_2TiS_4$ the metal displacements from ideal sites are noticeable. Mössbauer spectra reveal different magnetic behaviors; in $FeTi_2S_4$ there is a transition from paramagnetic to magnetic ordering at temperature $Tc$ = 145 K, giving rise to unusually low hyperfine magnetic field of 2.5 T (at 77 K) if compared with values of iron magnetic moments reported previously, this behavior is explained on the basis of blocking Fe localized magnetic moments by the Spin density wave (SDW) originated from 3d Ti atoms. In $Fe_2TiS_4$ a transition from paramagnetic to (SDW) arises at $Tc$ = 290 K, the SDW is spread in both Fe and Ti metals through [101] crystallographic plane and undergoes a transition of first order from Incommensurate SDW (ISWD) to Commensurate SDW (CSDW) at $T_{IC}$ = 255 K. The atomic positions in the unit cell are correlated to the magnetic behavior in both sulfides.

## 1. Introduction

Iron – titanium sulfides $FeTi_2S_4$ and $Fe_2TiS_4$ have monoclinic $Cr_3S_4$ - type structure as shown in Fig. 1. This structure is a metal-deficient NiAs-like one, in which there are two crystallographic sites for cations, one site in a metal-deficient layer ($M_I$) with *ordered metal vacancies* and the second site in a metal full layer ($M_{II}$). These layers are intercalated between hexagonal close-packed sulfur layers. The lattice parameters of $Cr_3S_4$ are a = 5.964 Å, b = 3.428 Å, c = 11.272 Å and $\beta$ = 91.50 $^o$ [1]. From X-ray studies about site distributions of Fe and Ti atoms it was *proposed* that in $FeTi_2S_4$, Fe atoms are located in $M_I$ layer and Ti atoms in $M_{II}$ layer [2], and in $Fe_2TiS_4$, the $M_I$ layer is occupied by Fe atoms while $M_{II}$ layer is occupied by Fe and Ti atoms with equal proportions [3].

$FeTi_2S_4$ is antiferromagnetic as reported from magnetic susceptibility measurements, with Neel temperature and paramagnetic moment for Fe atoms $T_N$ = 138 K, $\mu$/Fe = 3.51

$\mu_B$ [4], $T_N$ = 140 K, $\mu$/Fe = 3.40 $\mu_B$ [5] and $T_N$ = 140 K, $\mu_{eff}$/Fe = 3.60 $\mu_B$ [6] provided that Ti atom had no magnetic moment.

$Fe_2TiS_4$ is ferrimagnetic with Curie temperature *Tc* = 285 K as suggested from magnetic measurements using a torsion balance magnetometer [7].

In the literature there are three $^{57}$Fe Mössbauer studies on these iron titanium sulfides; two studies on $FeTi_2S_4$ and one study on $Fe_2TiS_4$.

The two Mössbauer studies on $FeTi_2S_4$ were made at variable temperature (4.2 K – 300 K) [8, 9], from which, we emphasize the significant result concerning the *small* hyperfine magnetic field (HF) at 4.2 K; HF ~ 3 T, which is in contradiction with the value of magnetic moments from magnetic susceptibility data mentioned above as stated by the authors.

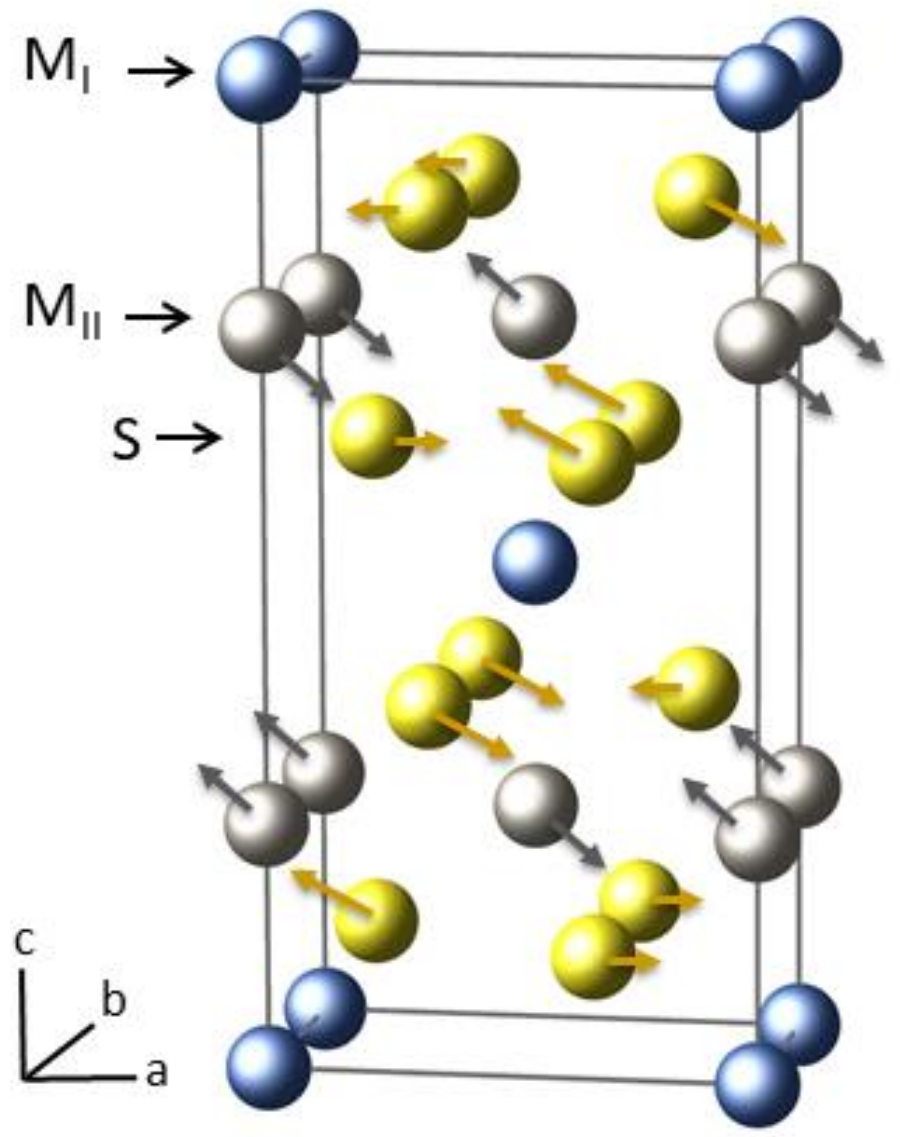

**Fig. 1**: The ideal structure of $Cr_3S_4$ (Space group I2/m, No. 12) after Jellinek [1], blue spheres: metals in the metal deficient layer ($M_I$), grey spheres: metals in the metal full layer ($M_{II}$) and yellow spheres: sulfur atoms. The arrows indicate directions and magnitudes (five times enlarged) of the observed deviations from the ideal atomic positions, see text.

The Mössbauer study on $Fe_2TiS_4$ is reported with spectra at two distinct temperatures; 291 K and 80 K [7], the spectra were analyzed as due to *one* site for Fe atoms even though these atoms occupy two inequivalent sites in the crystal structure as reported in X-ray studies [3], that is, the spectrum At 291 K, is non-magnetic (doublet), noticeably non-symmetrical, and at 80 K the spectrum is magnetic, fitted with the average hyperfine magnetic field <HF> = 19.8 T and broadened by the distribution of HF, the author stated that the estimation of the individual parameters for each one of the two iron sites was not possible.

In a previous paper, we reported a $^{57}$Fe Mössbauer study at variable temperature on related isostructural phases of iron vanadium sulfides $FeV_2S_4$ and $Fe_{1.8}V_{1.2}S_4$, in which the presence of antiferromagnetic itinerant magnetism in the form of Incommensurate spin density waves was proposed (ISDW) [10]. The ISDW behavior is revealed both in the shape of Mössbauer spectra and the metal displacements from ideal sites of $Cr_3S_4$ type structure, determined with Rietveld method.

Since the iron titanium sulfides $FeTi_2S_4$, $Fe_2TiS_4$ and iron vanadium sulfides $FeV_2S_4$ and $Fe_{1.8}V_{1.2}S_4$ are isostructural, and the fact that vanadium and titanium are close neighbors in the periodic table, it is a matter of interest to study the nature of the magnetism in the iron titanium sulfides in the same manner as done in iron vanadium sulfides.

The purpose of this investigation is to carry out a systematic study by $^{57}$Fe Mössbauer Spectroscopy at variable temperature on the iron titanium sulfides $FeTi_2S_4$ and $Fe_2TiS_4$. Additionally, a determination of atomic positions using Rietveld refinement method will also be reported, because unlike iron vanadium sulfides, no Rietveld study was reported to determine the refined atomic positions for $FeTi_2S_4$ and $Fe_2TiS_4$ at present.

## 2. Experimental

Samples were synthesized by direct reaction of elements following a procedure reported in the literature [11]; that is, Fe (99.999%), Ti (99.99%) and S (99.999%) (All from STREM CHEMICALS) were mixed in stoichiometric proportions and sealed in evacuated quartz tube at pressure of $10^{-3}$ Pa, then treated with the following sequence: (a) Heating to 773 K with rate 7 K / h, keeping at 773 K for one week, cooling until room temperature at natural rate i.e., shutting down the heater. (b) Repeating the step (a) with final temperature of 1173 K. (c) The samples were ground to secure more homogeneity and sealed again in an evacuated quartz tube at pressure of $10^{-3}$ Pa. (e) Repeating the step (a) with final temperature 1223 K and keeping this temperature for two weeks.

Powder X-Ray diffraction analysis was carried out using Philips X'pert diffractometer with Cok$\alpha$ radiation and Bruker D8 Advance diffractometer with Cuk$\alpha$ radiation. For Rietveld refinements of the crystal structure, the software GSAS was used [12], the refinement parameters are the Atomic positions ($x$, $y$, $z$), Site Occupation Factor ($N$) and Isotropic Temperature Factor ($U_{iso}$ in $\text{Å}^2$)

Figures of crystallographic phases were made using the software ‘‘Balls and Sticks’’ [13].

$^{57}$Fe Mössbauer spectra were recorded using a spectrometer running in a triangular symmetric mode. The driving unit used was made by Wissel GmbH, Germany, with $^{57}$Co source in Rh matrix. The liquid nitrogen cryostat, sample holder and temperature controller designed and built at Universidad Central de Venezuela (UCV). Data was acquired using a system made by Ortec Inc. TN, USA, and analyzed by curve fitting software developed in collaboration between Commissariat à l'Energie Atomique (CEA), Saclay, France and UCV, Venezuela. The calculated parameters are the Center Shift ($CS$ in mm/s) calibrated with metallic iron, the Quadrupole Splitting ($QS$ in mm/s), the Hyperfine Magnetic Field (HF in T), the angle between the Electric Field Gradient (EFG) and the HF ($\theta$) and the half width at half maximum of the absorption lines ($\Gamma$ in mm/s) assuming Lorentzian shape.

## 3. Results:

### *3.1. X-ray Diffraction*

Both $FeTi_2S_4$ and $Fe_2TiS_4$ were obtained successfully with no extra peaks than those indexed for monoclinic structure. Fig. 2 shows the XRD patterns treated with Rietveld refinement method and Table 1 shows the parameters of the unit cell together with those reported previously for purpose of comparison. Table 2 shows the other refined parameters *x*, *y*, *z*, *N* and $U_{iso}$ in addition to those *proposed* in previous reports for $Cr_3S_4$ and $FeTi_2S_4$. We can note that for $FeTi_2S_4$ the refined atomic positions are similar for those of the "ideal" structure, on the other side; for $Fe_2TiS_4$ the metals in $M_{II}$ layer are noticeably deviated from the ideal positions of *x* and *z* axes, Jellinek [1] reported a deviation of x and z parameters for metals in $M_{II}$ layer and sulfurs from their "ideal" values for the $Cr_3S_4$ phase - see Fig. 1 and Table 2-. It is interesting that these deviations are present in the monoclinic system $(Fe,V)_3S_4$ for *all* phases reported elsewhere; $V_3S_4$ [16,17], $FeV_2S_4$ [18-21] and $Fe_2VS_4$ [18,19], where the deviation in (x) for metals in $M_{II}$ layer is in the range of (x ~ -0.041 – -0.066) that is almost four and five times the deviation for Cr atoms in $Cr_3S_4$ (x ~ -0.012). These metal displacements were correlated to the itinerant magnetism in the isostructural $FeV_2S_4$ and $Fe_{1.8}V_{1.2}S_4$ sulfides [10], and as we will see in the discussion below, they play an important role in the magnetic properties of $FeTi_2S_4$ and $Fe_2TiS_4$.

### *3.2. $^{57}$Fe Mössbauer Spectroscopy*

#### *3.2.1. Spectra at room temperature*

Spectra at room temperature (RT) are shown in Fig. 3 and Mössbauer parameters in Table 3 together with those reported previously. The spectrum of $FeTi_2S_4$ is composed of one nonmagnetic sub spectrum (doublet) that is related to one crystallographic site for Fe atoms in $M_I$ layer, the values of *CS* and *QS* are in agreement with those reported previously [8,9]. The spectrum of $Fe_2TiS_4$ is composed of two doublets that are related to two different crystallographic sites, that is, one site in $M_I$ layer and another site in $M_{II}$ layer which is in agreement with Rietveld refinement results and those of previous X ray study [3]. This is the first Mössbauer evidence related to the existence of two sites of Fe atoms since the paper of Muranaka [7] in which the spectrum reported was unresolved for two Fe sites.

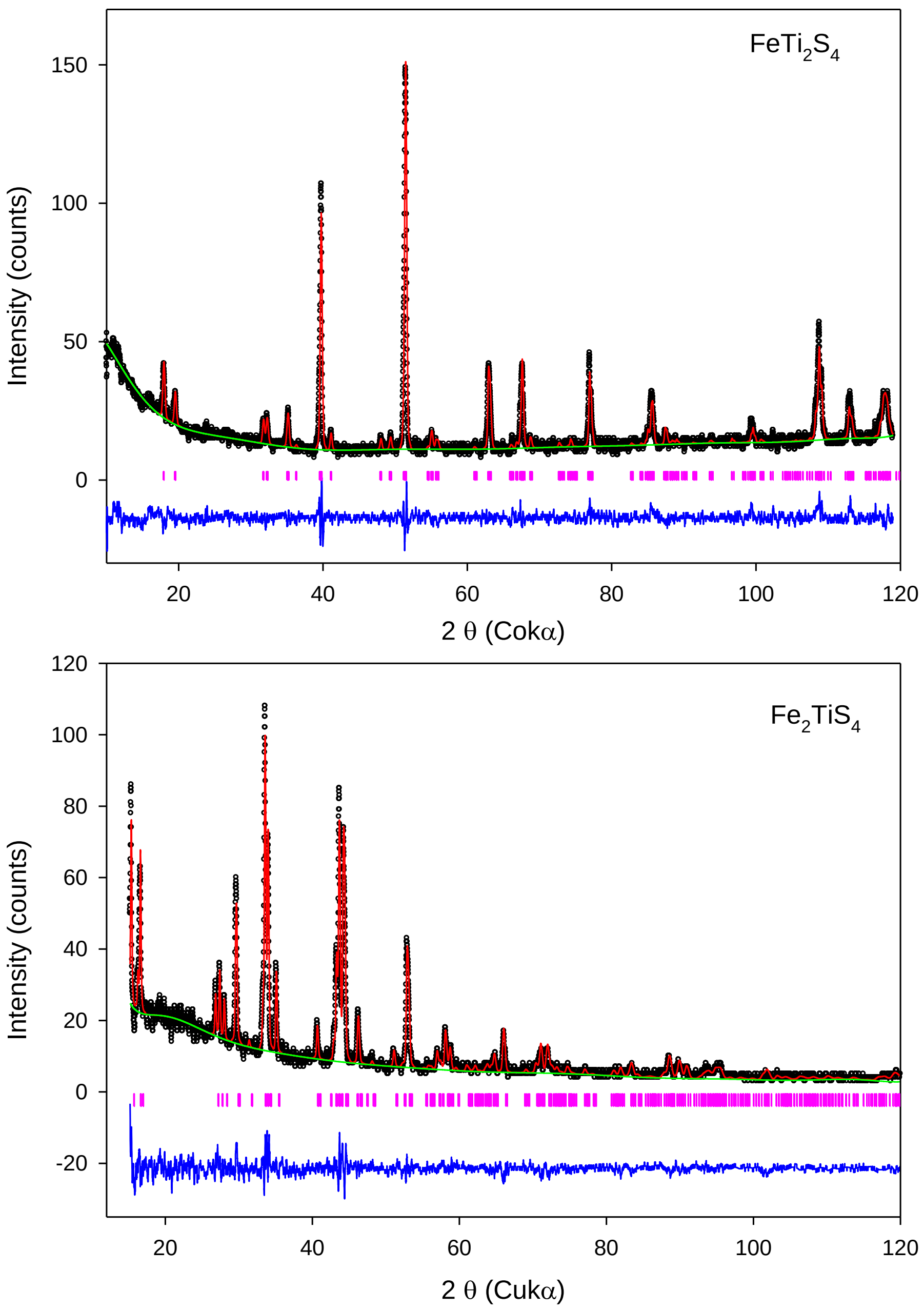

**Fig. 2:** Powder X ray patterns of $FeTi_2S_4$ (up) and $Fe_2TiS_4$ (down), symbols and lines represent the following: circles: experimental pattern, red line: calculated pattern, blue line: difference, green lines: background curve and pink markers: Bragg reflections.

**Table 1**: The Unit cell parameters of $FeTi_2S_4$ and $Fe_2TiS_4$, the uncertainties are 0.001 Å.

| Phase | a (Å) | b (Å) | c (Å) | $\beta$ (°) | Reference. |
|---|---|---|---|---|---|
| $FeTi_2S_4$ | 5.929 | 3.426 | 11.460 | 90.10 | [14] |
| $FeTi_2S_4$ | 5.929 | 3.426 | 11.465 | 90.10 | [3] |
| $FeTi_2S_4$ | 5.927 | 3.428 | 11.458 | 90.10 | [11] |
| $FeTi_2S_4$ | 5.953 | 3.437 | 11.498 | 90.10 | [2] |
| $FeTi_2S_4$ | 5.950 | 3.417 | 11.53 | 90.20 | [15] |
| $FeTi_2S_4$ | 5.950 | 3.421 | 11.497 | 89.87 | [6] |
| $FeTi_2S_4$ | 5.949 | 3.418 | 11.509 | 90.22 | [This study] |
| $Fe_2TiS_4$ | 5.979 | 3.432 | 11.161 | 91.70 | [3] |
| $Fe_2TiS_4$ | 5.980 | 3.430 | 11.160 | 91.70 | [7] |
| $Fe_2TiS_4$ | 5.969 | 3.426 | 11.269 | 91.38 | [This study] |

**Table 2**: Refined parameters of $FeTi_2S_4$ and $Fe_2TiS_4$, space group I2/m, No. 12.

| Sample / Reference. | atom | *x* | *y* | *z* | *N* | $U_{iso}$ (Å$^2$) |
|---|---|---|---|---|---|---|
| $Cr_3S_4$ (ideal) [1] | Cr1 | 0 | 0 | 0 | 1 | --- |
| | Cr2 | 0 | 0 | 1/4 | 1 | --- |
| | S1 | 1/3 | 0 | 3/8 | 1 | --- |
| | S2 | 1/3 | 0 | 7/8 | 1 | --- |
| $Cr_3S_4$ (real) [1] | Cr1 | 0 | 0 | 0 | 1 | --- |
| | Cr2 | -0.012 | 0 | 0.263 | 1 | --- |
| | S1 | 0.355 | 0 | 0.365 | 1 | --- |
| | S2 | 0.320 | 0 | 0.876 | 1 | --- |
| $FeTi_2S_4$ [2] | Fe1 | 0 | 0 | 0 | 1 | --- |
| | Ti2 | 0 | 0 | 1/4 | 1 | --- |
| | S1 | 1/3 | 0 | 3/8 | 1 | --- |
| | S2 | 1/3 | 0 | 7/8 | 1 | --- |
| $FeTi_2S_4$ [This study] | Fe1 | 0 | 0 | 0 | 1 | 0,030 |
| | Ti2 | -0.006 | 0 | 0.253 | 1 | 0.009 |
| | S1 | 0.328 | 0 | 0.374 | 1 | 0.002 |
| | S2 | 0.337 | 0 | 0.884 | 1 | 0.002 |
| $Fe_2TiS_4$ [This study] | Fe1 | 0 | 0 | 0 | 1 | 0.035 |
| | Fe2 | -0.031 | 0 | 0.260 | 0.5 | 0.008 |
| | Ti2 | -0.031 | 0 | 0.260 | 0.5 | 0.008 |
| | S1 | 0.332 | 0 | 0.364 | 1 | 0.002 |
| | S2 | 0.345 | 0 | 0.881 | 1 | 0.002 |

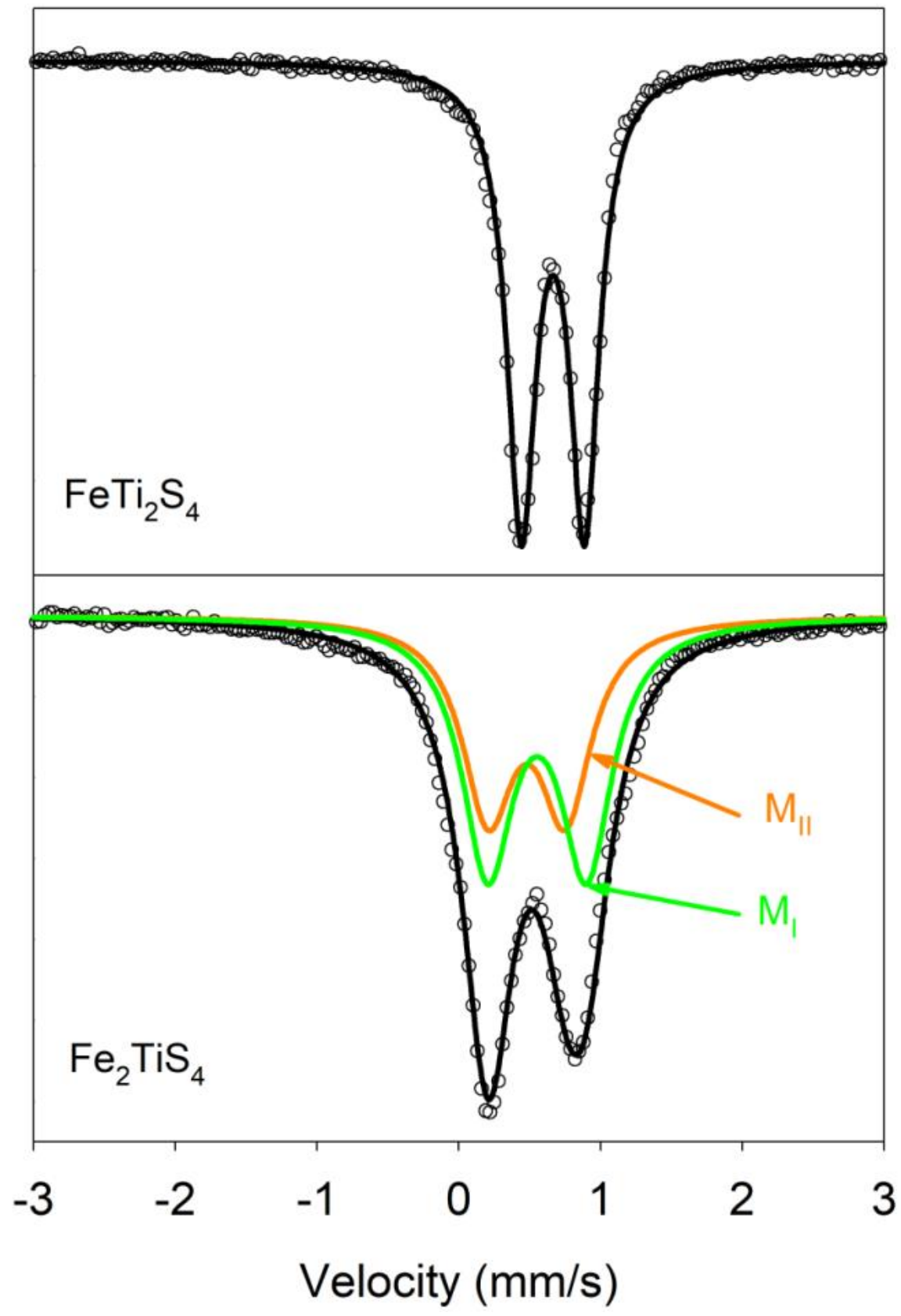

**Fig. 3:** Mössbauer spectra at RT.

**Table 3**: Parameters derived from Mössbauer spectra at room temperature [a].

| Sample / Reference | Fe site | *CS* (mm/s) ± 0.01 [a] | *QS* (mm/s) ± 0.01 [a] | Relative area (%) |
|---|---|---|---|---|
| $FeTi_2S_4$ [8] | $M_I$ | 0.78 | 0.44 | 100 |
| $FeTi_2S_4$ [9] | $M_I$ | 0.79 | 0.44 | 100 |
| $FeTi_2S_4$ [this study] | $M_I$ | 0.78 | 0.45 | 100 |
| $Fe_2TiS_4$ [7] | --- | 0.63 [b] | 0.52 [b] | 100 [b] |
| $Fe_2TiS_4$ [this study] | $M_I$<br>$M_{II}$ | 0.66<br>0.59 | 0.70<br>0.55 | 52<br>48 |

[a]) The uncertainties are in the parameters from this study.
[b]) Spectrum unresolved for two sites of iron atoms, see introduction section.

### *3.2.2. Spectra at variable temperature*

Selected spectra at variable temperature of $FeTi_2S_4$ and $Fe_2TiS_4$ are shown in Figs. 4 and 5 respectively and Table 4 shows the parameters derived from Mössbauer spectra at variable temperature. Due to the broad nature of the line absorption spectra, they were fitted with Hyperfine Field Distribution (HFD), unlike previous Mössbauer studies [7–9] where, in the case of $FeTi_2S_4$, the spectrum was fitted with simple 8 line Zeeman transition [8] or not fitted at all [9], and in the case of $Fe_2TiS_4$ the spectrum was also fitted with simple 8 line Zeeman transition [7]. From the Figs. 4 and 5 we note that the spectrum of $FeTi_2S_4$ is far from the usual six resolved Zeeman lines at all temperatures, while that of $Fe_2TiS_4$ becomes more resolved at lower temperatures. Due to the complexity of the spectra, it is convenient to describe their features and the parameters separately as follows:

1. Hyperfine Field Distribution (HFD)

The spectrum of $FeTi_2S_4$ is fitted with one HFD (right side of Fig. 4) which is related to one site for Fe atoms in $M_I$ layer, and that of $Fe_2TiS_4$ is fitted with two HFDs related to the sites of Fe atoms in $M_I$ and $M_{II}$ layers, the best fit is achieved when both HFDs have the same histogram (right side of Fig. 5).

2. Center Shift (*CS*) and Debye temperature ($\theta_D$)

The thermal evolution of *CS*, - see Fig. 6 - is fitted with Debye model [22]

$$CS(T) = IS + \delta_{SOD}(T) \qquad 1$$

with

$$\delta_{SOD}(T) = -\frac{9k_BT}{2Mc}\left[\frac{\theta_D}{8T} + \left(\frac{T}{\theta_D}\right)^3 \int_0^{\theta_D/T} \frac{x^3dx}{e^x - 1}\right] \qquad 2$$

where $IS$ is the intrinsic Isomer Shift, considered to be constant and $\delta_{SOD}(T)$ is the Second Order Doppler Shift, $M$ is the isotope mass, c is the speed of light and $\theta_D$ is the Debye Temperature. The fitting procedure yields to the values of $\theta_D$ = 250 K for Fe atoms in $M_I$ layer (both $FeTi_2S_4$ and $Fe_2TiS_4$), and $\theta_D$ = 550 K for Fe atoms in $M_{II}$ layer ($Fe_2TiS_4$). It is to be expected that the $\theta_D$ value in the $M_I$ layer is lower than that in the $M_{II}$ due to vacancy ordering around Fe atoms in $M_I$ layer. Similar behavior is present in

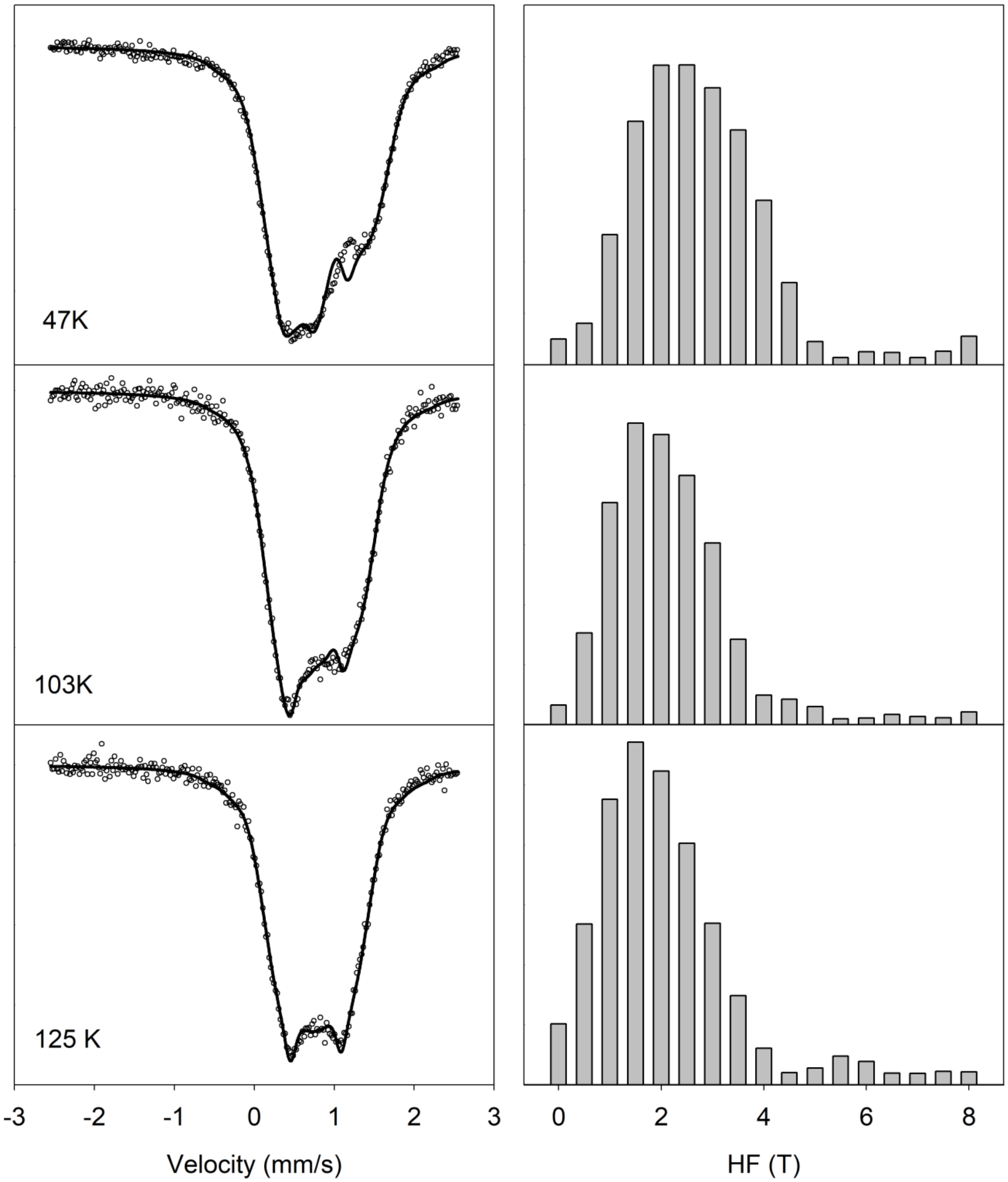

**Fig. 4**: Selected Mössbauer spectra at variable temperature of $FeTi_2S_4$ (left) with corresponding histograms (right).

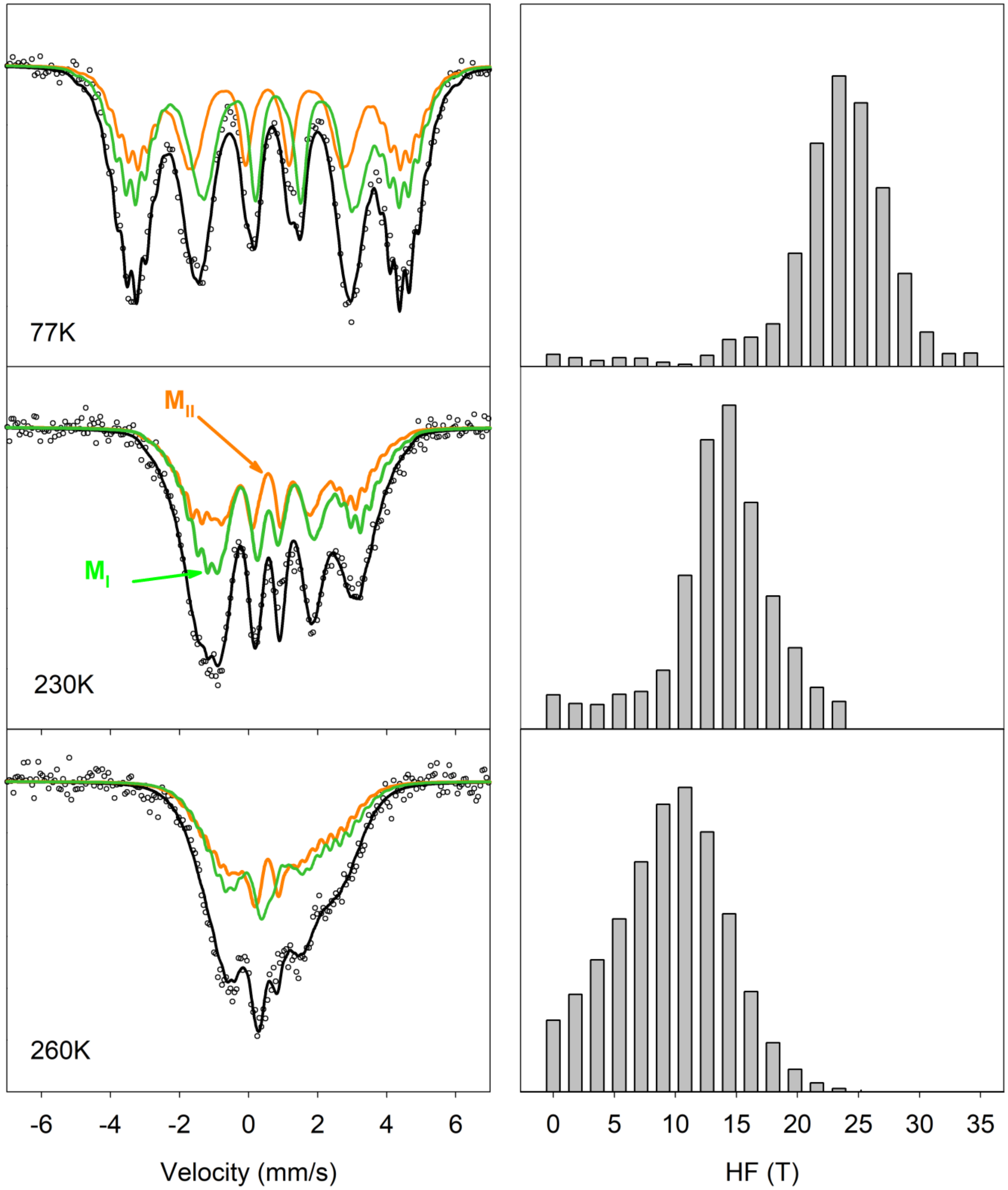

**Fig. 5**: Selected Mössbauer spectra at variable temperature of $Fe_2TiS_4$ (left) with corresponding histograms (right).

**Table 4**: Parameters derived from Mössbauer spectra at variable temperature. *CS*, *QS*, <HF> and θ are reported at 77 K. The parameters from references are reported at specified temperature between parentheses.

| Sample / Reference | Fe site | *CS* (mm/s) ± 0.01 | *QS* (mm/s) ± 0.01 | $\theta_D$ (K) ± 10 | *Tc* (K) ± 5 | <HF> (T) ± 0.2 | θ (°) ± 2 |
|---|---|---|---|---|---|---|---|
| $FeTi_2S_4$ [8] | $M_I$ | 0.94 (4.2 K) | --- | --- | 155 | 2.3 (4.2 K) | 37 [a] (4.2 K) |
| $FeTi_2S_4$ [9] | $M_I$ | 0.91 (150 K) | 0.56 (150 K) | --- | 140 | 3.0 (4.2 K) | --- |
| $FeTi_2S_4$ [this study] | $M_I$ | 0.93 | -0.64 | 250 | 145 | 2.5 | 68 |
| $Fe_2TiS_4$ [7] | --- | 0.67[b] | 0.58[b] | --- | 276[b] | 19.8[b] | --- |
| $Fe_2TiS_4$ [this study] | $M_I$<br>$M_{II}$ | 0.81<br>0.68 | -0.71<br>0.55 | 250<br>550 | 290 | 23.6<br>(80 K) | 40<br>30 |

[a]) θ = 65° at 130K.
[b]) Spectrum unresolved for two sites of iron atoms, see introduction section.

$FeV_2S_4$ and $Fe_{1.8}V_{1.2}S_4$ [10] although in the case of the present study the $\theta_D$ value in $M_I$ layer is even lower, i.e, 250 K in $FeTi_2S_4$ vs 380 K in $FeV_2S_4$.

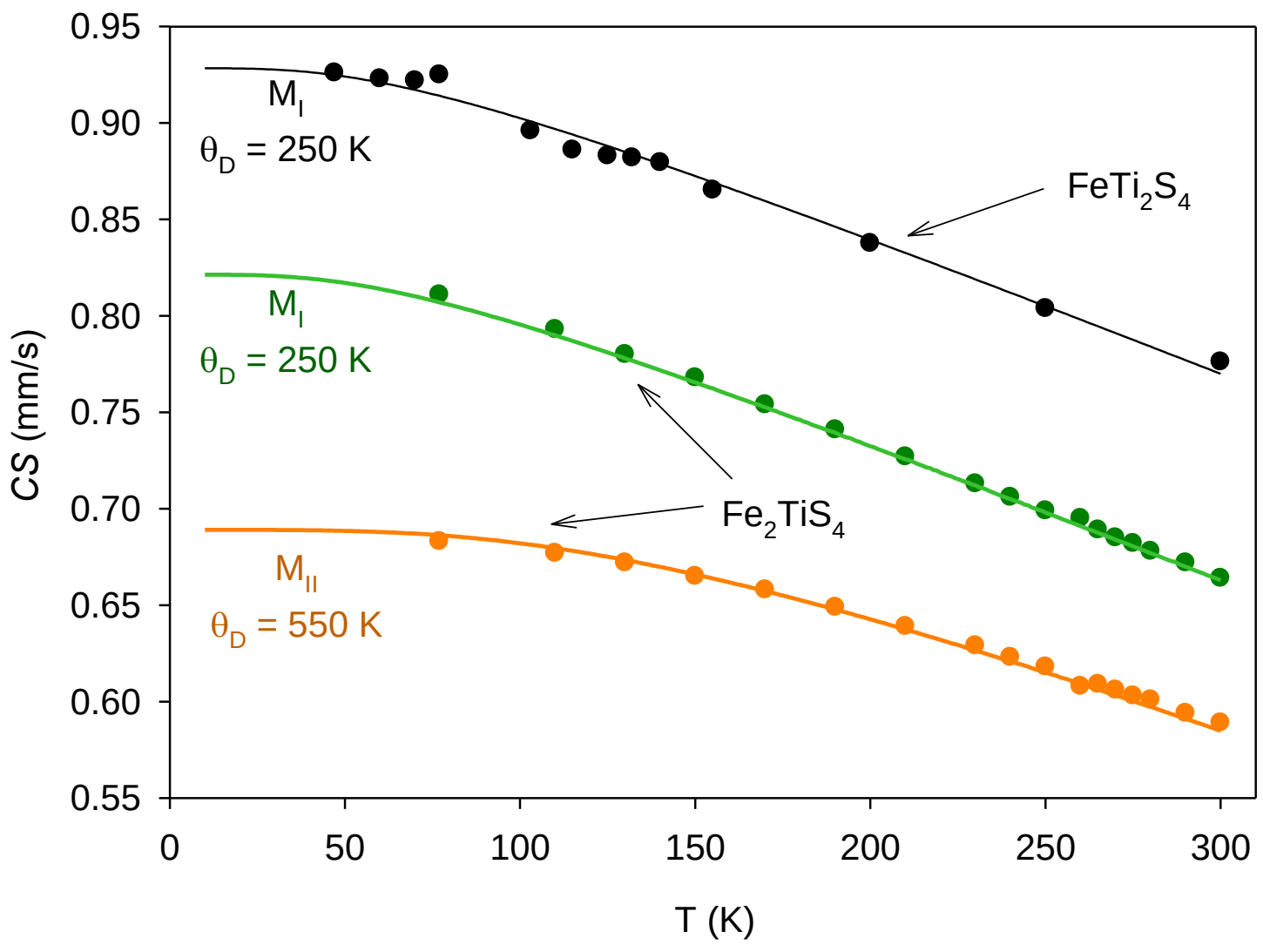

**Fig. 6:** Behavior of *CS* with temperature, circles denote the experimental results and lines denote the Debye model.

3. Quadrupole Splitting (*QS*)

In $FeTi_2S_4$, *QS* increases with decreasing temperature in absolute value, - see Fig. 7 - in agreement with previous Mössbauer studies [8, 9], the sign of *QS* is negative and is stabilized below transition temperature (*Tc)* where a magnetic ordering appears. In $Fe_2TiS_4$, *QS* does not change with temperature in either site of Fe atoms, and *QS* in the site $M_I$ is also negative below *Tc*. Similar behavior is reported in $FeV_2S_4$ and $Fe_{1.8}V_{1.2}S_4$ [10].

4. Average Hyperfine Field (<HF>)

The average HF, (<HF>) at 77 K has a value of 2.5 T in $FeTi_2S_4$ and 23.5 T in $Fe_2TiS_4$. The former value is comparable to that reported previously [8, 9] and the latter is slightly higher than that reported [7] (see the Introduction section). Also, it is interesting to note that in the $Fe_2TiS_4$ the value of <HF> is the same for both sites of Fe atoms in $M_I$ and $M_{II}$ layers because of the same HFD in both sites as mentioned above.

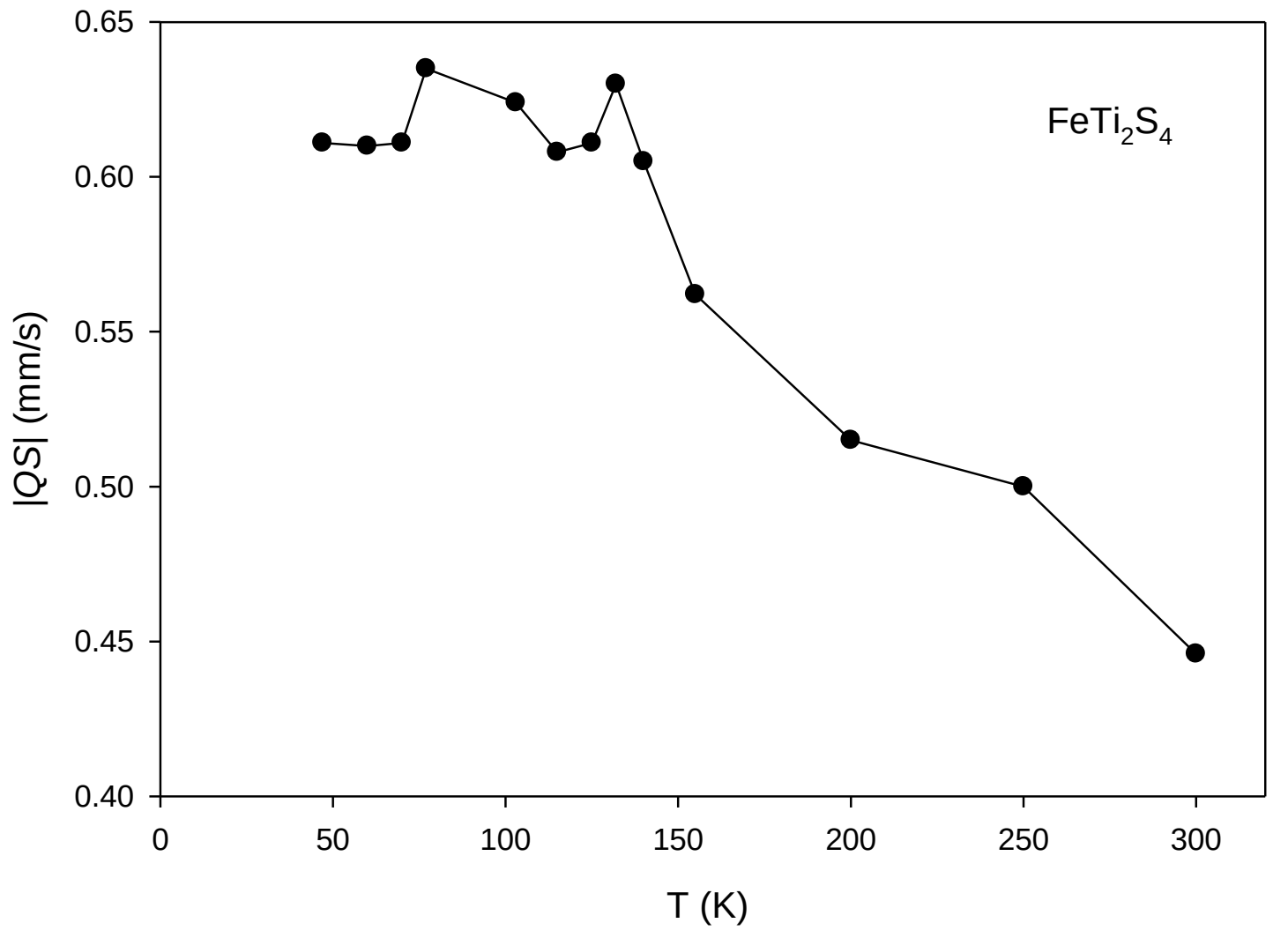

**Fig. 7:** Behavior of |*QS*| with temperature in $FeTi_2S_4$ sample, the sign of *QS* is negative, see text.

5. Transition temperature (*Tc*)

The estimated transition temperature (*Tc*) in $FeTi_2S_4$ is *Tc* = 145 ± 5 K, which is in agreement with the values derived from Mössbauer measurements [8, 9] and the values of $T_N$ derived from magnetic susceptibility data mentioned in the introduction section. In $Fe_2TiS_4$, *Tc* = 290 ± 5 K which differs from the value *Tc* = 276 obtained from a previous Mössbauer study but in agreement with the value of *Tc* = 285 K obtained from magnetic measurements using torsion balance magnetometer [7].

6. Angle between the Electric Field Gradient and the Hyperfine Magnetic Field (θ)

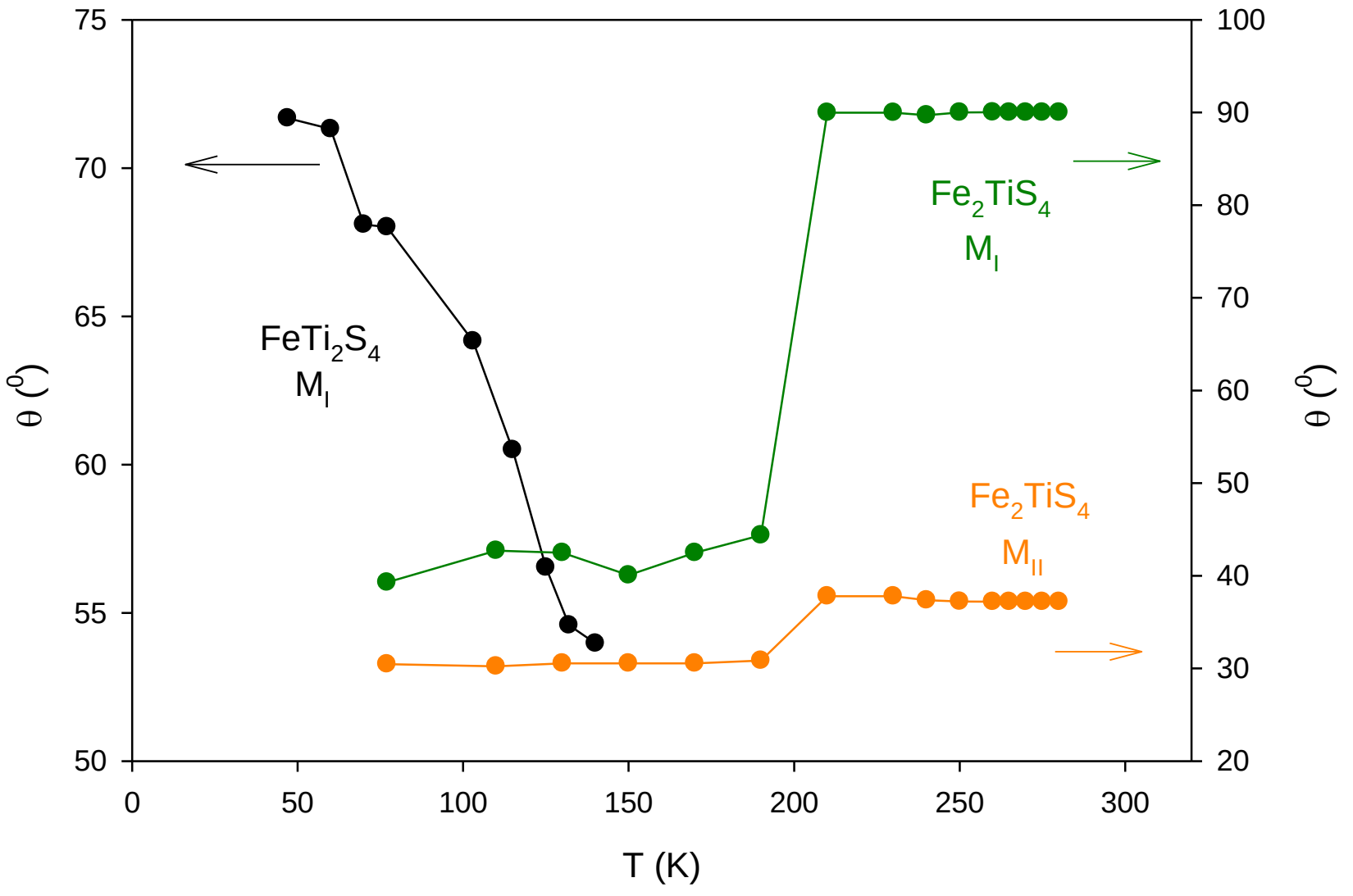

**Fig. 8:** Behavior of θ with temperature in $FeTi_2S_4$ and $Fe_2TiS_4$.

The behavior of the polar angle (θ) is shown in Fig. 8. We note in $FeTi_2S_4$ that θ increases with decreasing temperature from θ = 54 $^o$ (T = 140 K) to θ = 68 $^o$ (T = 77 K), and in $Fe_2TiS_4$, $\theta_I$ and $\theta_{II}$ (referred to $M_I$ and $M_{II}$ respectively) remain almost constant in two temperature ranges with values $\theta_I$ ~ 40 $^o$ and $\theta_{II}$ ~ 30 $^o$ in temperature range (77 – 190 K) and change abruptly to $\theta_I$ ~ 90 $^o$ and $\theta_{II}$ ~ 37 $^o$ in temperature range (210 – 290 K).

## 4. Discussion:

### *4.1. Atomic arrangements and conductivity*

$FeTi_2S_4$ has the atomic positions that are almost close to those of the ideal $Cr_3S_4$ type structure, while in $Fe_2TiS_4$ the metals in $M_{II}$ layer have noticeable displacements in x and z coordinates. In Fig. 9 are illustrated both the unit cell and the crystalline lattice of $FeTi_2S_4$ and $Fe_2TiS_4$. Applying the same notation used for the monoclinic $(Fe, V)_3S_4$ system we denote for the separation between metals in $M_I$ and $M_{II}$ layers as $d_{12}$ and the separation between metals in the same $M_{II}$ layer as $d_{22}$. The displacements of metals in $Fe_2TiS_4$ cause the formation of the staircase-like sheets with separation between them, which is denoted by $d_{ss}$. Table (5) shows the values of $d_{12}$, $d_{22}$ and $d_{ss}$. Note that when $d_{ss} \approx d_{22}$ the staircase-like sheets disappear which is almost the case of $FeTi_2S_4$.

In studies regarding transport properties of the $FeTi_2S_4$ sulfide [23, 24] the authors applied the Goodenough's rules [25, 26] for the critical separation between Ti atoms (Rc) for metallic conductivity in $FeTi_2S_4$ phase, which is estimated as Rc = 3.52 Å. So, in $FeTi_2S_4$, the $M_{II}$ layer should be metallic because the separations in this layer are b = 3.418 Å, $d_{22}$ = 3.366 Å and $d_{ss}$ = 3.497 Å. Applying the same argument for the $Fe_2TiS_4$, we can note that the crystalline lattice is formed by staircase-like sheets with a metallic conductivity along $d_{12}$ and $d_{22}$ chains and these sheets are separated by $d_{ss}$ = 3.778 Å which is higher than Rc. Below we will correlate this result to magnetic properties.

### *4.2. Magnetic properties;*

#### *4.2.1. $FeTi_2S_4$: Blocked Fe localized moments by Spin density wave of Ti atoms.*

The characteristic of low HF and the apparent contradiction with the value of local magnetic moments of Fe atoms determined by magnetic susceptibility data (see the introduction section) will be discussed. Fatseas [8] inferred that further studies might be undertaken to determine if possible field compensation by opposite field contribution may exist, and Katada [9] suggested that the small hyperfine magnetic field may be due to the anisotropy of the magnetic field at Fe nucleus with opposite field contributions.

In this study we propose that the origin of this opposite field is the Spin density wave (SDW) formed from Ti atoms in $M_{II}$ layer. The following indications support this hypothesis of the SDW compensation:

a) The Ti – Ti separations in $M_{II}$ layer favor the formation of SDW as they are lower than the critical separation for metallic conductivity Rc, and therefore the 3d orbitals are overlapped.

b) The value of *CS* (0.78 mm/s) is related to high spin $Fe^{2+}$, hence, bearing a localized magnetic moment, this result is the same as reported by Fatseas [8] and Katada [9] and the interpretation from these publications seems to be conflicting; that is; Fatseas [8] suggested that the measure of *CS* must be proceeded by a very careful calibration because the $FeTi_2S_4$ phase has a metallic type conductivity attributed to direct cation-cation interactions from the Goodenough critical distance Rc, Katada [9] suggested that the value of *CS* is consistent with the fact that the Fe-Fe bonding distance is too large for a direct metal atom interaction and thus it is assumed that the Fe ion has only localized d electrons. In this paper we propose that the interpretations offered by these two authors are *not conflicting but complementary,* because the metallicity suggested by the former author is due to Ti – Ti interaction, that is, SDW in $M_{II}$ layer, while local magnetic moment suggested by the latter author is attributed to isolated Fe atoms in the vacancy layer $M_I$. Also the two studies suggested a kind of compensation mechanism that is responsible of lowering the HF of Fe atoms.

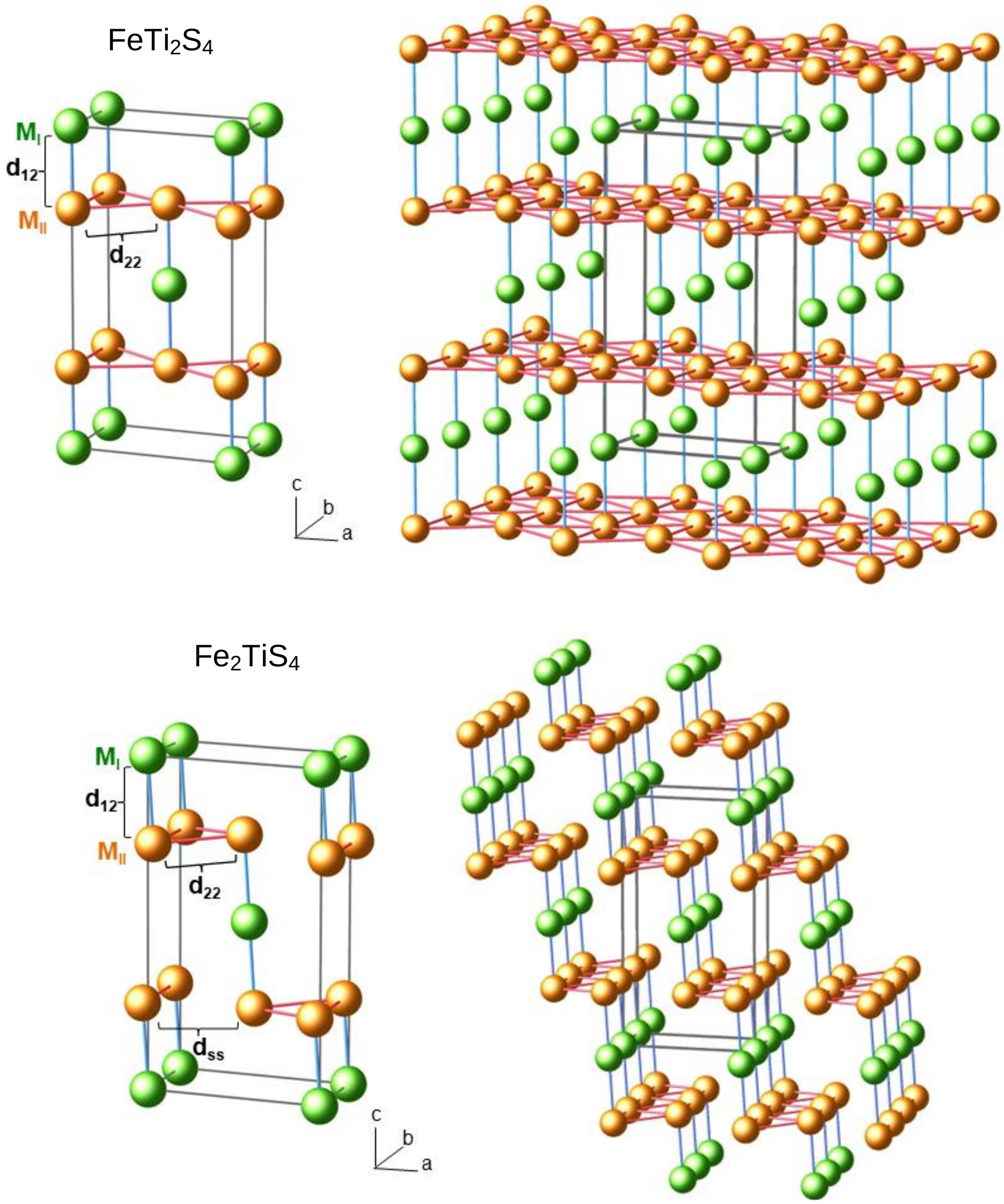

**Fig. 9:** Atomic arrangements in $FeTi_2S_4$ (up) and $Fe_2TiS_4$ (down). Left: The unit cell where distances between metals in $M_I$ and $M_{II}$ layers are denoted as $d_{12}$ and distances between metals in the same $M_{II}$ are denoted as $d_{22}$. Right: Case of $FeTi_2S_4$: The crystalline lattice where $M_I$ and $M_{II}$ layers are alternated periodically. Right: Case of $Fe_2TiS_4$: The crystalline lattice formed by staircase-like sheets where each sheet is composed by the two metallic bonds $d_{12}$ and $d_{22}$. The separation between sheets is denoted $d_{ss}$. Sulfur atoms are omitted for the sake of simplicity.

c) The Fe – Ti separation $d_{12}$ (see Fig. 9 and Table 5) favors the interaction between the localized magnetic moments of Fe in $M_I$ layer and the 3d conduction electrons – SDW – of the Ti atoms in $M_{II}$ layer, another indication of this interaction is the behavior of resistivity with temperature, where a minimum is reported [6].

The mechanism of compensated Fe localized moments by SDW is present in isostructural phase $FeTi_2Se_4$ where Mössbauer spectrum at room temperature is a paramagnetic doublet with *CS* = 0.79 mm/s and the HF at 4.2 K was estimated about (4T), which is in apparent contradiction with the value of magnetic moment, $\mu_{Fe} = 4.0\ \mu_B$, determined from magnetic susceptibility [27]. The authors gave no explanation of the low HF but compared it to the value reported for the $FeT_{i2}S_4$ [8]. Another report with Neutron Diffraction study on $FeTi_2Se_4$ [28] suggested the formation of commensurate SDW from Ti conduction electron spins in $M_{II}$ layer that are strongly polarized by the Fe localized moments at $M_I$ layer.

The mechanism of compensated Fe localized moments by SDW was reported for the first time in Cr-Fe alloys with Fe concentration from 0.2 to 5 at. % studied with $^{57}Fe$ Mössbauer Spectroscopy [29], in which the low HF value (3.5 T at 4.2 K) was interpreted in terms of a spin-compensated magnetic moment of Fe atoms that was partially destroyed by the SDW exchange field of Cr atoms.

**Table 5**: Cation separations in $FeTi_2S_4$ and $Fe_2TiS_4$.

| Sample | $d_{12}$ (Å) | $d_{22}$ (Å) | $d_{ss}$(Å) |
|---|---|---|---|
| $FeTi_2S_4$ | 2.915 | 3.366 | 3.497 |
| $Fe_2TiS_4$ | 2.941 | 3.130 | 3.778 |

### 4.2.2. $Fe_2TiS_4$: Incommensurate and commensurate spin density wave.

Applying the same analysis made in (Fe, V)$_3$S$_4$ system [10] we suggest that the evolution of Mössbauer spectra are consistent with the presence of SDW due to the overlapping 3d orbitals both from Fe and Ti atoms along $d_{12}$ and $d_{22}$ bonds. The *CS* values in $M_I$ and $M_{II}$ layers (0.664 and 0.591 mm/s) support the evidence of the overlapping 3d orbitals. The SDW has a 2D character placed in the staircase-like sheets located in [101] crystallographic plane where the inter–sheet separation $d_{ss}$ = 3.778 Å (see Fig. 9). Note that Fe atoms in the $M_I$ and $M_{II}$ sites "see" the same HF, given that the spectra in both sites *have the same* HFD as mentioned in results section.

From the histograms of Fig. 5 we can estimate the magnetic and non-magnetic portions in the similar way did in (Fe, V)$_3$S$_4$ system. In the case of the $Fe_2TiS_4$, the non- magnetic portion is limited to 5 T. The evolution of magnetic portion with temperature is shown in Fig. 10. The magnetic portion is attributed to the condensate density of the conduction electrons in the SDW state, and the non-magnetic portion is attributed to the conduction electrons in the normal state, i.e., the non-condensate electrons. From Fig. 10 we note that with decreasing temperature, the SDW undergoes a transition from incommensurate spin density wave (ISDW) to commensurate spin density wave (CSDW) at $T_{IC} \approx 255$ K, and this transition is of first order. This ISDW-CSDW transition could be related to the behavior of θ with temperature - see Fig. 8 – and the behavior of spontaneous magnetization with temperature [7] where a sharp change arises at T ~ 260 K.

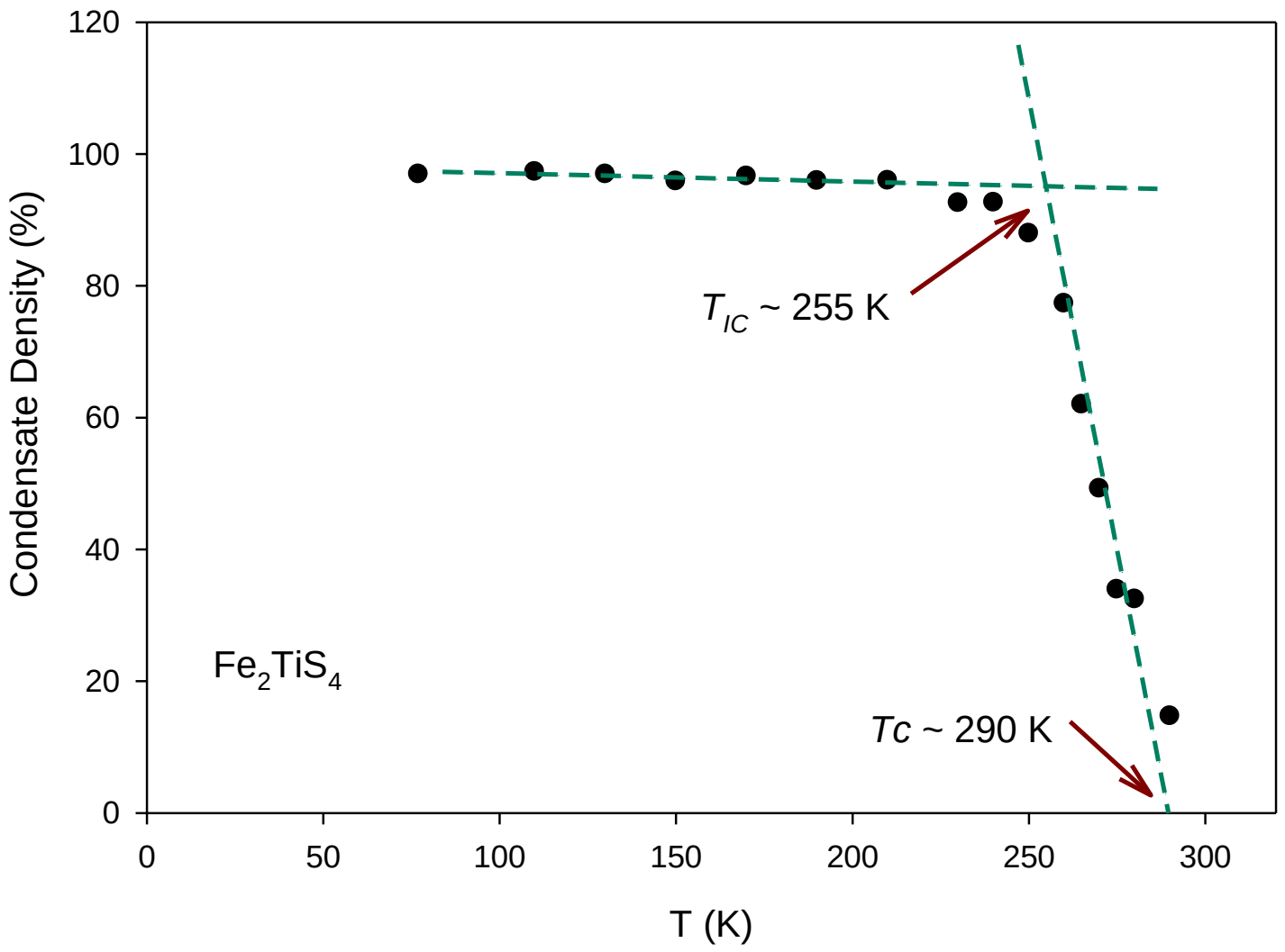

**Fig. 10:** Magnetic portion - Condensate density - vs temperature in $Fe_2TiS_4$. Dashed lines are guides for the eye.

## 5 Conclusions

We conclude that the magnetic properties of $FeTi_2S_4$ and $Fe_2TiS_4$ are affected by the atomic arrangements in the unit cell, which is a monoclinic $Cr_3S_4$ type one. In $FeTi_2S_4$, the atomic positions are almost identical to the ideal positions of $Cr_3S_4$ type structure. On the other hand, Mössbauer spectra are consistent with the existence of blocking mechanism over Fe local magnetic moments due to itinerant 3d electrons - SDW - of Ti atoms, causing therefore a low Hyperfine Magnetic Field (HF) reported in previous Mössbauer studies. In $Fe_2TiS_4$ there are atomic displacements from ideal $Cr_3S_4$ type structure that favor the formation of spin density wave (SDW) in the [1 0 1] crystallographic plane, and Mössbauer spectra evidence a transition from incommensurate SDW (ISDW) to commensurate SDW (CSDW) which is of first order.